\documentclass{iau} 
\usepackage{graphicx}

\newcommand{\Mbh}{M_{\rm BH}}
\newcommand{\Mout}{M_{\rm out}}
\newcommand{\Min}{M_{\rm in}}
\newcommand{\epsEM}{\epsilon_{\rm EM}}
\newcommand{\epsw}{\epsilon_{\rm w}}
\newcommand{\Mstar}{M_*}
\newcommand{\Msun}{M_{\odot}}
\newcommand{\Lx}{L_{\rm X}}
\newcommand{\Lbh}{L_{\rm BH}}
\newcommand{\Ledd}{L_{\rm Edd}}
\newcommand{\Medd}{M_{\rm Edd}}

\newcommand{\Lsn}{L_{\rm SN}}
\newcommand{\Rsn}{R_{\rm SN}}
\title[JD 11.~~AGN feedback and star formation in ETGs] %% give here short title %%
{AGN feedback and star formation in ETGs: negative and positive
feedback}

\author[Luca Ciotti et al.]   %% give here short author list %%
{Luca Ciotti$^1$, Jeremiah P. Ostriker$^{2,3}$, Andrea Negri$^{1,4}$,\\
Silvia  Pellegrini$^1$, Silvia Posacki$^1$, \and Greg Novak$^5$}

\affiliation{$^1$Dept. of Physics and Astronomy, University of
  Bologna, via Ranzani 1, I-40127 Bologna, Italy,
 \\ email: {\tt luca.ciotti@unibo.it} \\[\affilskip]
$^2$Dept. of Astrophysical Sciences, Princeton University, NJ, USA \\[\affilskip]
$^3$Dept. of Astronomy and Astrophysics, Columbia University, NY, USA \\[\affilskip]
$^4$ Institut d'Astrophysique de Paris, CNRS, UMR 7095 98bis bvd Arago,
F-75014 Paris, France\\[\affilskip]
$5$ Observatoire de Paris, LERMA, CNRS, 61 Av de l'Observatoire, F-75014 Paris, France}

\pubyear{2015}
\volume{315}  %% insert here IAU Symposium No.
\jname{From interstellar clouds to star-forming galaxies}
\editors{P. Jablonka, F. Van der Tak \& P. Andr\'e, eds.}
\begin{document}

\maketitle

\begin{abstract}

  AGN feedback from supermassive black holes (SMBHs) at the center of
  early type galaxies is commonly invoked as the explanation for the
  quenching of star formation in these systems. The situation is 
  complicated by the significant amount of mass injected in the
  galaxy by the evolving stellar population over cosmological
  times. In absence of feedback, this mass would lead to unobserved
  galactic cooling flows, and to SMBHs two orders of magnitude more
  massive than observed. By using high-resolution 2D hydrodynamical
  simulations with radiative transport and star formation in
  state-of-the-art galaxy models, we show how the intermittent AGN
  feedback is highly structured on spatial and temporal scales, and
  how its effects are not only negative (shutting down the recurrent
  cooling episodes of the ISM), but also positive, inducing star
  formation in the inner regions of the host galaxy.

\end{abstract}

\firstsection % if your document starts with a section,
\section{Introduction}

In a quite widespread view, merging is considered the main cause of
QSO activity, because it is believed to be, after the end of the
galaxy formation epoch, the only mechanism to add fresh gas to the
central SMBHs in early type galaxies (ETGs). However, it has also been
shown (e.g., Ciotti \& Ostriker 1997) that the stellar mass losses
produced during stellar evolution cyclically feed a central gas
inflow, and then trigger the QSO activity in isolated ETGs (sometimes
considered ``red and dead''). Indeed, recent observations indicate
that the QSO activity can be independent of galaxy merging.  QSO
activity is also invoked as the explanation of star formation
quenching in ETGs (negative feedback) but, as stressed in Ciotti \&
Ostriker (2007, hereafter CO07)), AGN feedback can induce star formation (positive
feedback) during ``cooling flow'' episodes fueled by stellar mass
losses. In fact, also in isolation, significant amounts of fresh gas
($\simeq 20-30\%$ of the initial stellar mass $\Mstar$ of the galaxy,
depending on the IMF) are injected over the galaxy body by stellar
evolution, at a rate approximately given by $\dot\Mstar (t)\propto
\Mstar (0)\, t^{-1.3}$ (Ciotti et al. 1991, see also Pellegrini
2012). These losses interact with the pre-existing hot ISM, and mix
with it, due to thermalization of the
stellar velocity dispersion. In a galaxy of total B-band luminosity
$L_{\rm B}$, SNIa's explosions at a rate $\Rsn (t)\propto L_{\rm B}\,
t^{-s}$ provide additional mass and heat to the ISM, with rates $\dot M_{\rm
  SN}(t)=1.4\Msun\,\Rsn (t)$ and $\Lsn(t) = 10^{51} {\rm erg}\,\Rsn
(t)$. Since recent estimates suggest $s\simeq 1$, then the SNIa`s specific heating $\Lsn/{\dot\Mstar}$
increases with time. The resulting hot atmosphere however cannot be
static.  The interstellar mass keeps increasing, its cooling time
decreases, until a cooling catastrophe takes place in the central
galactic region and the SMBH feedback takes place.

In a series of papers, Ciotti, Ostriker and co-workers, with the aid
of high-resolution hydrodynamical simulations, and a physically based
feedback implementation, studied in detail the AGN activity cycle
induced by accretion of stellar mass losses, also considering star
formation. In particular, it was shown that AGN feedback is able to
avoid the overgrowth of the SMBH that would be caused by accretion of
the mass lost by stars if uninpeded, and also that the same feedback
can induce significant star formation at the center of ETGs (CO07). By
using state-of-the-art galaxy dynamical models, and an updated 
version of the 2D code of Novak et al. (2011, 2012, hereafter N12,
and references therein), we are now investigating in detail the ISM
behavior in realistic ETG models with evolving input ingredients from
the stellar population (Ciotti et al., in preparation).

\section{The models}

We consider axisymmetric two-component galaxy models with adjustable
flattening embedded in a spherically symmetric NFW dark matter halo
and with a central SMBH. The stellar component is described by the
ellipsoidal deprojection of the de Vaucouleurs law.  The total
gravitational field, the solution of the two-integrals Jeans
equations, and the projection of the resulting kinematical field on
the plane of the sky are computed as described elsewhere (Posacki et
al. 2013).  The velocity dispersion and rotational fields of the
stellar populations, needed to compute energy and momentum injection
in the ISM due to stellar evolution, are computed by using a
generalized Satoh $k$-decomposition (Ciotti \& Pellegrini 1996).  The
E4 and E7 galaxy models are obtained by flattening (at fixed stellar
mass $\Mstar$) spherical ``progenitors'' with central (aperture)
velocity dispersion of 180, 210, 250 and 300 km/s, and 
reproduce observed scaling laws (Negri et al. 2014). The initial mass of the SMBH is taken to be
$\Mbh=10^{-3}\Mstar$, near to the Magorrian relation, as expected at
the end of the period of galaxy formation.

The hydrodynamical equations and the input physics are given in Ciotti
\& Ostriker (2012), and integrated by an improved version of the N12
code.  Mass, momentum and energy sources and sinks associated with
evolution of the passively evolving initial stellar population, and
with the new stars added by star formation, are treated as described
in Negri et al. (2015, hereafter N15). In particular the star
formation rate is given by $\dot{\rho}_{\rm SF} = 0.1\rho
/\max(\tau_{\rm cool}, \tau_{\rm dyn})$, where $\rho$ is the local ISM
density, and $\tau_{\rm cool}$ and $\tau_{\rm dyn}=\min(\tau_{\rm
  Jeans}, \tau_{\rm rot})$ are the cooling and dynamical times,
respectively.  In N15 we showed that this prescription reproduces the
observed Kennicutt-Schmidt law remarkably well.  Radiative and
mechanical feedback from the accreting SMBH (both depending on the
hydrodynamically self-consistently determined mass accretion rate on
the SMBH, $\dot\Mbh$), are finally considered.  In the energy
equation, photoionization and Compton heating and cooling ($T_{\rm C}
\simeq 2\times 10^7$ K), bremsstrahlung and line cooling are taken
into account, with modifications due to the ionization effects of AGN
radiation. In the momentum equation, radiation pressure due to AGN
activity is computed by solving the radiative transport (in spherical
symmetry) for the accretion luminosity $\Lbh$, considering
photoionization+Compton opacity, and electron scattering. For
simplicity at this stage we ignore the detailed treatment of dust
formation and destruction, and the associated radiation pressure
effects (Hansley et al. 2014, N12), as well as the
radiation pressure due to star ligh in different bands (e.g., see
Ciotti \& Ostriker 2012).  For given $\dot\Mbh$, then, one has:
\begin{equation}
  \Lbh = \epsEM\dot\Mbh c^2, \quad\quad
\epsEM = {\epsilon_0 A \dot{m}\over 1 + A \dot{m}},\quad\quad 
\dot{m}\equiv{\dot\Mbh\over \dot\Medd} = 
{\epsilon_0 \dot\Mbh c^2\over \Ledd},
  \label{eq:lum}
\end{equation}
where $A=100$, $\epsilon_0=0.125$ and $\Ledd$ is the Eddington 
luminosity.

%%%%%%%%%%%%%%%%%%%%%%%%%%%%%%%%
\begin{figure}[b]
\vspace*{-1.0 cm}
\begin{center}
 \includegraphics[width=5.0in]{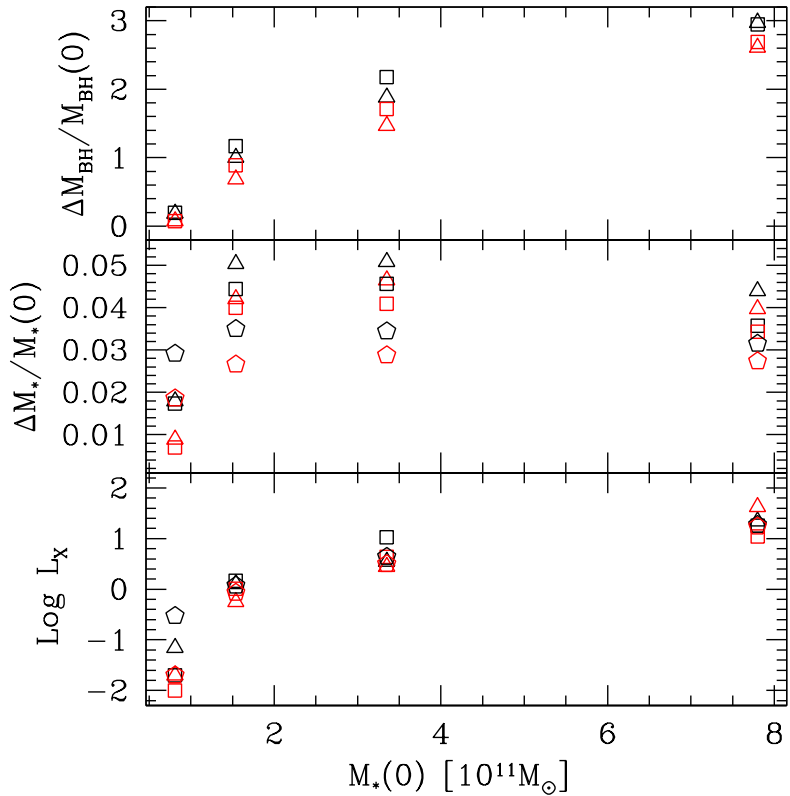} 
\vspace*{-1.0 cm}
 \caption{Global properties of the models at the end of the 
   simulations (13 Gyr), as a function of their initial stellar mass, for E4 
   (black) and E7 (red) models.  Triangles, squares, and pentagons 
   refer to models with radiative plus mechanical AGN feedback,
   mechanical AGN feedback only, and no AGN feedback, respectively.}
   \label{fig1}
\end{center}
\end{figure}
%%%%%%%%%%%%%%%%%%%%%%%%%%%%%%%%

Mechanical feedback is due to mass, momentum, and energy injection of
a conical-shaped nuclear wind, with half-opening angle of $\simeq
45^\circ$.  In terms of solid angle, this means that the wind is
visible from $\sim 1/4$ of the available viewing angles. The relevant
relations here are $\dot\Mbh = \dot\Min/(1+\eta)$, $\dot\Mout = \eta
\dot\Mbh$, $ \dot{p}_{\rm w} = \dot\Mout v_{\rm w}$, $ L_{\rm w} =
\epsw\dot\Mbh c^2$, where $\eta \equiv 2\epsw c^2 / v_{\rm w}^2$,
$\dot\Min$ and $\dot\Mout$ are the mass inflow and outflow rates at
the first grid point (a few pc from the SMBH). In the current
simulations, $\epsw$ follows a prescription similar to that adopted
for $\epsEM$, with maximum value of $10^{-4}$, while the wind velocity
$v_{\rm w}$ is $10^4$ km/s.

\section{Results}

As illustrated in Fig. 1 (top panel), the central SMBH grows more, in
absolute terms ($\Delta\Mbh$) and also in percentual ($\Delta\Mbh/\Mbh
(0)$), for increasing stellar mass $\Mstar$ of the host galaxy; the
growth reaches a factor of $\simeq 3$ in the biggest galaxies with AGN
feedback (to be compared with a factor $\simeq 30$ in models without
AGN feedback). Moreover, at any fixed $\Mstar$, more mass is accreted
by the SMBH in E4 galaxies than in E7 galaxies. This is due to the
fact that edge-on flattened galaxies are less bound (see N15) than
more spherical systems. As expected, at fixed $\Mstar$ and galaxy
shape, the accreted mass increases from models with full AGN feedback
(triangles), to models with mechanical AGN feedback only (squares),
and finally to models without AGN feedback (pentagons). Therefore,
since we start with a SMBH mass $\sim$ on the Magorrian relation, this
relation is quite well preserved thanks to the effect of AGN feedback,
which is able, in combination with large-scale SNIa heating, to
maintain small the SMBH masses, avoiding the accretion of $\simeq
99\%$ of the mass injected by the stars over the galaxy body.

The stellar mass added to the galaxy due to star formation in the gas
supplied 
by evolving stars ($\Delta\Mstar$, middle panel), is
larger in E4 galaxies than in E7 galaxies of same $\Mstar$. The total
mass in new stars is of the order of $4-5\%$ of the initial $\Mstar$,
indicating that an important fraction of the injected mass ($\simeq
0.2-0.3\Mstar$) escapes as a galactic wind, supported by SNIa heating
with the additional contributions of the thermalization of stellar
winds, of type SNII explosions in the new stellar populations, and of
the AGN feedback.  In general, as already found in the case of
spherical models (CO07), models with AGN feedback
tend to form {\it more} stars than models without (with the exception
of very low mass galaxies, where AGN feedback is able to sustain
global galactic winds).  This positive feedback is explained by
considering that the AGN prevents the rapid fall of the cooling
material to the center, allowing for more time for star formation. A
clear sign of this effect is the characteristic size of the
region of star formation, which is confined to the very center of
the galaxy in absence of AGN feedback (a small fraction of the optical
effective radius), while in models with AGN feedback it reaches a size of
the order of a kpc or more. The results of this work (in absence of
AGN feedback) agree well with those extensively discussed in N15
including star formation.

An important observational feature of the models is of course the
evolution of their luminosity, both of the central accreting SMBH
($\Lbh$), and of the galaxy hot ISM ($\Lx$). In Fig. 1 we show $\Lx$
in the X-ray $0.3-8$ keV band at the end of the simulations, with
values in nice agreement with observations, and only moderately
dependent on AGN feedback. The evolution of $\Lbh$ (not shown), is
confirmed to be highly fluctuating, as already found in our previous
works based on simpler galaxy models. The duty-cycle
(phenomenologically defined as the fraction of time spent by the SMBH
at $\Lbh$ higher than a few percents of $\Ledd$), is of the order of
$\simeq 3-5\%$.

\end{document}